\documentclass[preprintnumbers,10pt,nofootinbib]{revtex4}

\usepackage{amsmath,latexsym,amssymb,amsfonts}
\usepackage[dvips]{graphicx}
\usepackage{bm}

\begin{document}

\preprint{YITP-14-68}

\title{Instantons near a tachyonic top in anti de Sitter and the no-boundary regulator}

\author{\textsc{Bum-Hoon Lee$^{1,2}$}\footnote{{\tt bhl{}@{}sogang.ac.kr}},
\textsc{Wonwoo Lee$^{1}$}\footnote{{\tt warrior{}@{}sogang.ac.kr}},
\textsc{Daeho Ro$^{1}$}\footnote{{\tt dhro{}@{}sogang.ac.kr}}
and \textsc{Dong-han Yeom$^{1,3,4}$}\footnote{{\tt innocent.yeom{}@{}gmail.com}}
}

\affiliation{$^{1}$\small{Center for Quantum Spacetime, Sogang University, Seoul 121-741, Republic of Korea}\\
$^{2}$\small{Asia Pacific Center for Theoretical Physics, Pohang 790-784, Republic of Korea}\\
$^{3}$\small{Yukawa Institute for Theoretical Physics, Kyoto University, Kyoto 606-8502, Japan}\\
$^{4}$\small{Leung Center for Cosmology and Particle Astrophysics, National Taiwan University, Taipei 10617, Taiwan}}

\begin{abstract}
We investigate instantons near a tachyonic top in an anti de Sitter (AdS) background.
If the mass scale around the hill-top is above the Breitenlohner-Freedman (BF) bound, then the top is classically stable. When the BF bound is satisfied, it is already known that there can exist instantons with a non-zero probability, though fine-tunings of the potential are required. On the other hand, we may consider a possibility to obtain instantons with a non-zero probability for more natural shape of potentials. We found that the no-boundary regulator is useful to assign a non-zero probability for general instantons near the tachyonic top with a consistent framework. To use the no-boundary regulator, we need to introduce the complexification of fields. One interesting feature is that, for these AdS instantons, the classicality may not be satisfied after the Wick rotation and hence after the nucleation. This magnifies a novel difference between dS and AdS; a semi-classical boundary observer in AdS may notice the dispersion of quantum fields as a kind of uncertainty, while every semi-classical observers in dS are classicalized individually and hence there is no semi-classical observer who can see the quantum dispersion of the scalar field.
\end{abstract}

\maketitle

%\newpage

\tableofcontents

\newpage

\section{Introduction}

Since the foundation of AdS/CFT correspondence \cite{Maldacena:1997re}, various properties of the anti de Sitter (AdS) space become big interests of us. AdS has a gravitationally attractive property via a cosmological constant $\Lambda < 0$ and hence it gives somehow non-intuitive phenomena. For example, intuitively a hill-top of a tachyonic potential implies a unstable equilibrium and it might be destabilized by a small perturbation; on the other hand, because of the attractive property of the AdS space, a tachyonic top of AdS can be stable up to perturbations, as long as the mass scale $-m^{2} < 0$ of the tachyonic top is smaller than a critical value: $m^{2}/|\Lambda| < 3/2$. This condition is known by the Breitenlohner-Freedman (BF) bound \cite{Breitenlohner:1982jf}.

Then the next natural question is this: if the mass scale of the tachyonic top is smaller than the BF bound, then can the tachyonic top be destabilized via \textit{non-perturbative effects}? This question can be refined as follows:
\begin{itemize}
\item[1.] For which conditions, is the tachyonic top stable or unstable?
\item[2.] Is there a non-perturbative instanton around a tachyonic top? If there is, for which conditions?
\item[3.] If there is a non-perturbative instanton, does it mean the destabilization of the tachyonic top? If it does not mean the destabilization, then what does this mean?
\end{itemize}
For the first question, there is literature in terms of the energy of the AdS background \cite{Boucher:1984yx}. If the trivial AdS is the absolute minimum of possible energy variations, then for any modification of the configuration space, the AdS background should be stable. Although we cannot classify all possible stable potentials, at least we can give a definite example. For the second question, we know that there exists a potential that allows a non-perturbative and non-trivial instanton solution around the tachyonic top \cite{Kanno:2012ht}. However, it is still unclear whether the instanton means a destabilization of the tachyonic top or not.

Regarding the third question, our strategy is \textit{to compare with dS instantons}. There is a deep relation between a tachyonic top of AdS and a stable local minimum of dS \cite{Hertog:2011ky}. More specifically, for a tachyonic potential in AdS with $U(\phi)$,
\begin{eqnarray}
U(\phi) &\rightarrow& V(\phi) = - U(\phi), \nonumber \\ 
\alpha &\rightarrow& a = -i\alpha \nonumber 
\end{eqnarray}
connect the model from tachyonic AdS to stable dS, where $\alpha$ is the scale factor of AdS and $a$ is the scale factor of dS. We will discuss more details and justifications in the following sections. Anyway, this correspondence between AdS and dS implies that, \textit{if there is an instanton in dS, then there should be a corresponding instanton in AdS}. One more interesting correspondence is that if the mass scale becomes sufficiently larger than the BF bound, then there is no classicalized instanton around the local minimum of dS \cite{Hartle:2007gi}; while the tachyonic top is unstable up to small perturbations around the local maximum of AdS. So to speak, \textit{if there is no instanton in dS, then (corresponding) AdS is no more stable.} This is surely more than an accident.

Keeping this in mind, let us summarize the basic properties of instantons around the local minimum of dS \cite{Hartle:2007gi,Hwang:2011mp,Hwang:2012mf}.
\begin{itemize}
\item[A.] Instantons are basically \textit{fuzzy}, in the sense that solutions are complex-valued (the only real-valued instanton is the Hawking-Moss instanton around the local minimum or maximum \cite{Hawking:1981fz}). Of course, after the Wick rotation, the solutions approach real values in order to satisfy classicality \cite{Hartle:2007gi}.
\item[B.] Instantons are \textit{continuous} around the local minimum \cite{Hartle:2007gi,Hwang:2012mf}. This gives a positive dispersion of the scalar field around the local minimum \cite{Hwang:2012mf}, where this is consistent with a different approach \cite{Linde:1993xx}.
\item[C.] We interpret that this dispersion is due to the \textit{thermal/quantum fluctuations}. The positive dispersion is a kind of stationary field distribution via quantum fluctuations \cite{Hwang:2012mf,Linde:1993xx}. However, this does \textit{not} correspond to the bounce solution, such as Coleman-DeLuccia instantons \cite{Coleman:1980aw}, that destabilizes the local minimum.
\item[D.] This dispersion can be observed by only the omnipotent observer who can see all histories in superspace, while semi-classical observers cannot see the dispersion of the field \cite{Hwang:2012mf}.
\end{itemize}
So, if there is a correspondence between AdS and dS instantons, then can we say that (A) AdS instantons are complex-valued, (B) instantons are continuous around the tachyonic top, (C) they do not destabilize the tachyonic top, and (D) no observer can see the positive dispersion? In conclusion, we will say yes for (A), (B), (C), while say no for (D). Of course, we will refine and clarify the statements in the following sections.

In SEC.~\ref{sec:qua}, we discuss motivations regarding this problem. For many cases, instantons around the tachyonic top will have diverging action, while this seems to be strange in some points. First, we clarify this problem. In SEC.~\ref{sec:nob}, we discuss on the no-boundary regulator to deal with this problem. The introduction of the no-boundary regulator indeed covers various issues those were discussed in this introduction. Finally, in SEC.~\ref{sec:dis}, we summarize our results and discuss on future directions.

\section{\label{sec:qua}Quantum instability of hill-top in AdS}

Let us first investigate Euclidean instantons of a tachyonic hill-top potential in AdS space. We can show that there exist instantons like thick-shell, but their Euclidean actions diverge except the case when we finely tune the potential. We briefly review on this issue.

\subsection{\label{sec:insta}Instantons near the hill-top in AdS}

We consider the following model in Euclidean signatures:
\begin{eqnarray}
S_{\mathrm{E}} = -\int d^{4}x \sqrt{+g} \left( \frac{1}{16\pi}R - \frac{1}{2} (\nabla \phi)^{2} - U(\phi) \right),
\end{eqnarray}
where $R$ is the Ricci scalar, $\phi$ is a scalar field, $U(\phi)$ is a potential of $\phi$ such as
\begin{eqnarray}
U(\phi) = U_{0} - \frac{1}{2} m^{2} \phi^{2},
\end{eqnarray}
with $U_{0} < 0$. This describes a tachyonic top in AdS.

To investigate instantons, we impose $O(4)$ symmetric metric ansatz:
\begin{eqnarray}\label{eq:mini}
ds_{\mathrm{E}}^{2} = \frac{d\tau^{2} + \alpha^{2}(\tau) d\Omega_{3}^{2}}{|U_{0}|},
\end{eqnarray}
where $d\Omega_{3}^{2} = d\chi^{2} + \sin^{2}\chi (d\theta^{2} + \sin^{2} \theta d\varphi^{2})$.
Then the equations of motion are as follows:
\begin{eqnarray}
\ddot{\phi} &=& - 3 \frac{\dot{\alpha}}{\alpha} \dot{\phi} - \mu^{2} \phi,\\
\ddot{\alpha} &=& - \frac{8 \pi}{3}\alpha \left( \dot{\phi}^{2} - 1 - \frac{1}{2}\mu^{2} \phi^{2} \right),
\end{eqnarray}
where $\mu^{2}\equiv m^{2}/|U_{0}|$. The initial conditions are
\begin{eqnarray}
\alpha(0) = 0, \;\;\; \dot{\alpha}(0) = 1, \;\;\; \dot{\phi}(0) = 1.
\end{eqnarray}

In this case, the decay probability is
\begin{eqnarray}
P \propto e^{- \Delta S_{\mathrm{E}}},
\end{eqnarray}
where $\Delta S_{\mathrm{E}} = S_{\mathrm{E}}(\mathrm{solution}) - S_{\mathrm{E}}(\mathrm{background})$. In general, the Euclidean actions for solution and background diverge in AdS, but the subtracted result can be finite. Let us check how to subtract and for which conditions the decay rate can have non-zero probability.

\begin{figure}
\begin{center}
\includegraphics[scale=0.5]{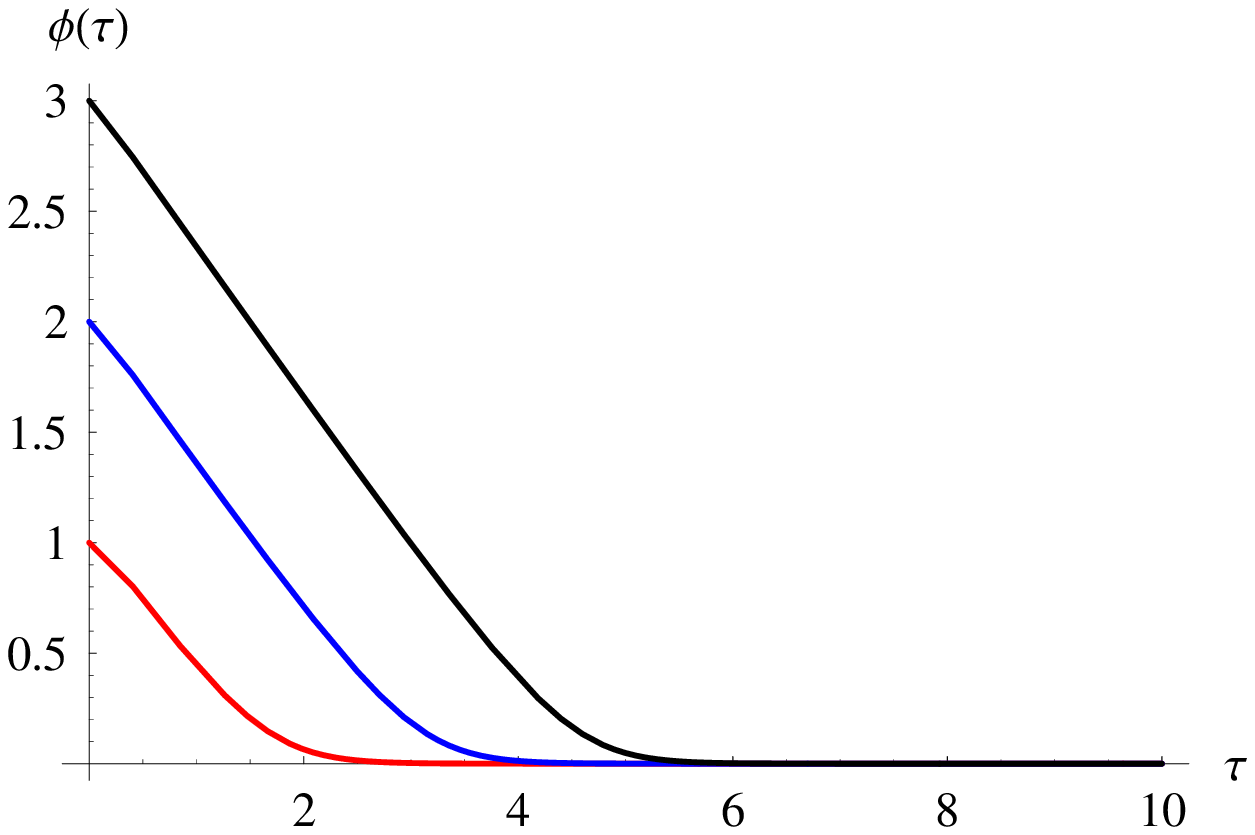}
\includegraphics[scale=0.5]{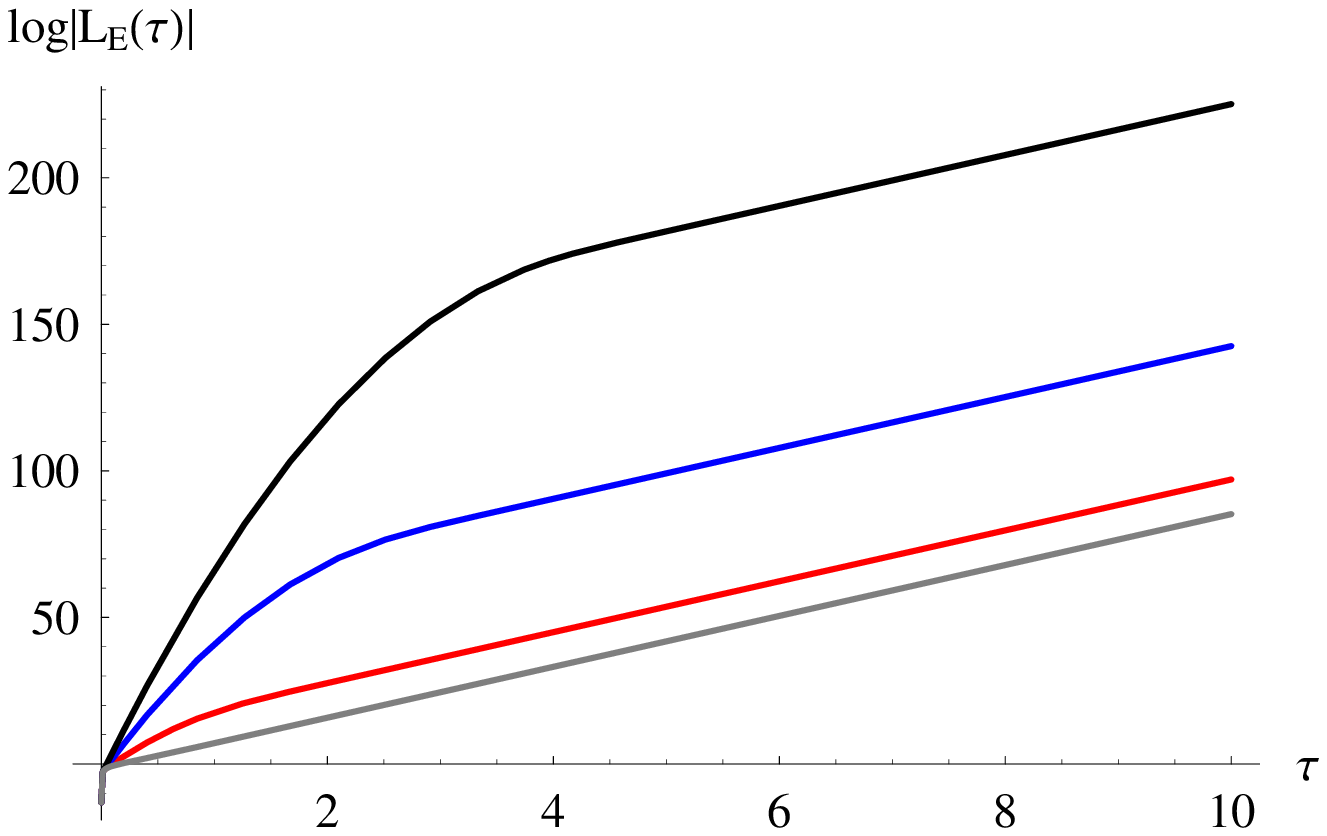}
\caption{\label{fig:phi2_2}Left: Some solutions for $V(\phi) = - 1 - \mu^{2} \phi^{2}/2$ with $\mu = \sqrt{6\pi} - 0.1$ ($\phi(0) = 3$ (black), $\phi(0)=2$ (blue), $\phi(0)=1$ (red)). Right: The Lagrangian $\log |L_{\mathrm{E}}|$ for each solutions and the background AdS solution (gray).}
\end{center}
\end{figure}
\begin{figure}
\begin{center}
\includegraphics[scale=0.5]{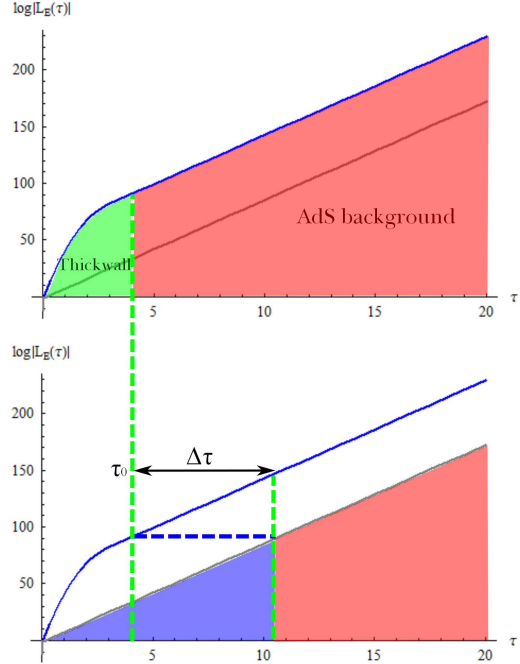}
\caption{\label{fig:concept}Upper: The Euclidean action of the instanton is the integral over the green colored region and the red colored region. Lower: The Euclidean action of the background is the integral over the blue colored region and the red colored region. Both of red colored region would be canceled by the subtraction.}
\end{center}
\end{figure}

FIG.~\ref{fig:phi2_2} demonstrates some numerical examples of this kind of solutions. As $\phi$ approaches to the top ($\phi \ll 1$), the solution approaches AdS and this phenomenon is universal, in the sense that the log of the Euclidean Lagrangian linearly increases with the same gradient. Then we can match the anti de Sitter background and try to subtract the background contributions, similar to the thin-shell cases.

FIG.~\ref{fig:concept} gives the conceptual picture. Upper means the Lagrangian of the Euclidean instanton solution. The Euclidean action is the integral over the green colored region and the red colored region (of course, the picture is a log of the Lagrangian and hence we have to consider the correct exponential factor), where the green colored region has $\phi \gtrsim 1$ and hence corresponds the thick-wall region and the red colored region has $\phi \ll 1$ and hence very similar as the AdS background. Lower means the Lagrangian of the pure AdS background. The Euclidean action is the integral over the blue colored region and the red colored region, where we choose the time slice between the blue colored region and the red colored region $\tau_{0} + \Delta \tau$ to have the same Lagrangian value.

Therefore, intuitively, the subtraction between the upper (instanton) and the lower (background) will give finite results because the red colored regions of upper and lower should be canceled. However, in terms of detailed calculations, this intuitive explanation does not hold in general. We see the details in the following subsections.

\subsection{\label{sec:condi}Conditions for non-zero probability}

In this subsection, we assume $U_{0}=-1$ without loss of generality. Around the local maximum, it is approximately pure Euclidean AdS such that
\begin{eqnarray}
\ddot{\alpha} \simeq + \frac{8\pi}{3} \alpha,
\end{eqnarray}
and hence
\begin{eqnarray}
\alpha \simeq \frac{1}{H_{0}} \sinh H_{0}\tau,
\end{eqnarray}
where $H_{0}^{2} = 8\pi /3$.

Now we can write the equation for the scalar field as follows:
\begin{eqnarray}
\ddot{\phi} + 3 H_{0} \dot{\phi} \coth H_{0}\tau + \mu^{2} \phi = 0.
\end{eqnarray}
For $H_{0}\tau \gg 1$, we further approximate
\begin{eqnarray}
\ddot{\phi} + 3 H_{0} \dot{\phi} + \mu^{2} \phi \simeq 0.
\end{eqnarray}
Therefore, if $\mu^{2}/H_{0}^{2} < 9/4$ (i.e., satisfies the BF bound), then
\begin{eqnarray}
\phi &\simeq& A_{1} \exp \left[ \left(- \frac{3}{2}H_{0} + \sqrt{\frac{9}{4}H^{2}_{0} - \mu^{2}}\right) \tau \right] + A_{2} \exp \left[ \left(- \frac{3}{2}H_{0} - \sqrt{\frac{9}{4}H^{2}_{0} - \mu^{2}}\right) \tau \right] \label{eq:noapp}\\
&\simeq& A_{1} \exp \left[ \left(- \frac{3}{2}H_{0} + \sqrt{\frac{9}{4}H^{2}_{0} - \mu^{2}}\right) \tau \right], \label{eq:app}
\end{eqnarray}
and hence approaches to local maximum without oscillation. Note that the approximation to Eq.~(\ref{eq:app}) is valid for the case when $A_{1}$ is not zero. Of course, $A_{1}$ can be zero, but it depends on the initial conditions. In our instanton solutions, we already fixed the initial conditions except the field amplitude. As the field approaches to the local maximum of the potential, the approaching speed can be controlled if we tune the shape of the potential; then by tuning the initial field value, we may be able to obtain suitable $A_{1}$. This means that unless we finely tune the shape of the potential, in general $A_{1}$ is non-zero, and hence the approximation to Eq.~(\ref{eq:app}) should be valid\footnote{Note that if we consider the tachyonic $\phi^{4}$ type potential \cite{Lee:2012ug}, near the local maximum the effective mass $\mu$ becomes zero. Therefore, if we can sure that the overall amplitude emerges to the local maximum as $\tau$ increases, then we can sure that $A_{1}$ is zero for this $\phi^{4}$ potential case. We investigate instantons around the $\phi^{4}$ potential in \cite{LLRY}.}.

Let us compare to the convergence condition of Kanno, Sasaki and Soda \cite{Kanno:2012ht}; we introduce their results without further derivation. In the paper, they used the conformal coordinate:
\begin{eqnarray}
ds_{\mathrm{E}}^{2} = \alpha^{2}(z) \left( dz^{2} + d\Omega_{3}^{2} \right).
\end{eqnarray}
Noted that, around the boundary ($z \rightarrow 0$ and $x \equiv \tanh z \rightarrow 0$), the solution should behave
\begin{eqnarray}
\alpha(x) &\simeq& \frac{1}{H_{0}} \frac{\sqrt{1-x^{2}}}{x} \times (1 - b x^{n}) + ... \simeq \frac{1}{H_{0}} x^{-1} + ..., \\
\phi(x) &\simeq& \frac{2}{\sqrt{8\pi}} \sqrt{\frac{2(n+1)}{n}b} x^{n/2} + ...,\\
U(\phi) &\simeq& - 1 + \frac{n(n-6)\pi}{3} \phi^{2} + ... \equiv  - 1 - \frac{1}{2} \mu^{2} \phi^{2} + ...,
\end{eqnarray}
where
\begin{eqnarray}
\mu^{2} = -\frac{2(n-3)^{2}\pi}{3} + m_{\mathrm{BF}}^{2} < m_{\mathrm{BF}}^{2}
\end{eqnarray}
and $m_{\mathrm{BF}} = \sqrt{6\pi}$ is the mass parameter at the BF bound. If we rewrite $n$, then
\begin{eqnarray}
\frac{n}{2} = \frac{3}{2} \pm \sqrt{\frac{9}{4} - \frac{\mu^{2}}{H_{0}^{2}}}.
\end{eqnarray}
In addition,
\begin{eqnarray}
\alpha \simeq \frac{1}{H_{0}x} \simeq \frac{1}{H_{0}}e^{H_{0}\tau}.
\end{eqnarray}
%\begin{eqnarray}
%a \simeq \frac{1}{H_{0}\sinh z} \simeq \frac{1}{H_{0}z} \simeq \frac{1}{H_{0}x} \simeq \frac{1}{H_{0}}e^{H_{0}\tau}.
%\end{eqnarray}
Therefore,
\begin{eqnarray}
\phi(\tau) \simeq A \exp{\left[\left(-\frac{3}{2} \mp \sqrt{\frac{9}{4} - \frac{\mu^{2}}{H_{0}^{2}}} \right)H_{0}\tau\right]} + ...
\end{eqnarray}
is obtained. Note that $n=2$ or $n>3$ is the condition that has a non-zero decay rate, while $n=2$ is unphysical in the sense that the energy diverges \cite{Kanno:2012ht}. If $n > 3$, then we should choose the upper sign, i.e., for the non-zero decay rate, the solution should follow the fast-decay mode.

To summarize, unless $A_{1}$ of Eq.~(\ref{eq:noapp}) is zero, the Euclidean action around the tachyonic top in AdS diverges. That is the reason why authors of \cite{Kanno:2012ht} had to tune the potential to obtain an instanton with a non-zero decay rate. Therefore, for the tachyonic $\phi^{2}$ potential, there is no Euclidean instanton that has a non-zero decay rate.

\subsection{Motivation of this paper}

Let us discuss motivations for further investigations.

First, there is a conceptual tension between our intuitive understanding of solutions in AdS space (SEC.~\ref{sec:insta}) and analytic calculations (SEC.~\ref{sec:condi}). This tension is related to the fact as follows: AdS is contracting in Lorentzian signatures and hence its physical volume is effectively bounded by $l \sim 1/\sqrt{|\Lambda|}$, while it has infinite volume in Euclidean signatures and has infinite entropy (or, entropy is not well-defined). On the other hand, dS is expanding in Lorentzian signatures and hence its physical volume is unbounded (again, volume of dS is not well-defined); while it has a finite volume (again, order of $1/\sqrt{\Lambda}$) in Euclidean signatures and hence has finite entropy. In this sense, if any quantum effect in AdS is disallowed because of very far effects (e.g., $r \gg l$), then it seems quite strange in terms of Lorentzian observer's point of view. For dS cases, the Euclidean treatment gives an effective cutoffs \cite{Hwang:2012mf} so that any observables to be finite. Can the Euclidean approach allow a similar treatment for AdS?

Second, we can find an example of potentials that the subtracted Euclidean action can be finite \cite{Kanno:2012ht}. However, if we slightly tilt the potential, then the slow-decay mode would be turned on and hence the subtracted action will diverge again. However, the real physics (existence of instantons) should not seriously affected by a very small tilt of the potential and hopefully we should explain their continuous variations.

These may indicate that there will be some treatment such that even though a solution does not satisfy the non-zero decay rate condition (of SEC.~\ref{sec:condi}), it still has physical meaning. In this paper, we assert that the no-boundary regulator, that was introduced by Hartle and Hertog \cite{Hertog:2011ky}, will do a proper role for this purpose.

\section{\label{sec:nob}No-boundary wave function and time contours}

In this section, we first reformulate Euclidean quantum gravity for dS background and second recover the conventions of the previous section to introduce the no-boundary regulator for AdS.

\subsection{No-boundary wave function}

The wave function of universe or the ground state wave function of the universe \cite{Hartle:1983ai} is
\begin{eqnarray}
\Psi\left[ h_{\mu\nu}, \psi \right] = \int \mathcal{D}g\mathcal{D}\phi\;\; e^{-S_{\mathrm{E}}\left[g_{\mu\nu}, \phi\right]},
\end{eqnarray}
where the Euclidean action is
\begin{eqnarray}\label{eq:action}
S_{\mathrm{E}} = -\int d^{4}x \sqrt{+g} \left( \frac{1}{16\pi}R - \frac{1}{2} (\nabla \phi)^{2} - V(\phi) \right).
\end{eqnarray}
Now we keep in mind the dS space with the potential
\begin{eqnarray}
V(\phi) = V_{0} + \frac{1}{2} m^{2} \phi^{2},
\end{eqnarray}
and we sum-over all regular and compact geometries those have the only boundary $\partial g_{\mu\nu} = h_{\mu\nu}$ and $\partial \phi = \psi$.

We choose the minisuperspace metric ansatz:
\begin{eqnarray}\label{eq:mini}
ds_{\mathrm{E}}^{2} = \frac{d\tau^{2} + a^{2}(\tau) d\Omega_{3}^{2}}{V_{0}}.
\end{eqnarray}
(Because of this metric choice, to investigate dynamics, we can choose $V_{0} = 1$ without loss of generality.) The Euclidean action reduces to
\begin{eqnarray}
S_{\mathrm{E}} = \frac{2 \pi^{2}}{V_{0}} \int d\tau \left[ -\frac{3}{8\pi} \left( a \dot{a}^{2} + a \right) + \frac{1}{2}a^{3} \dot{\phi}^{2} + a^{3} \left( 1 + \frac{1}{2} \mu^{2} \phi^{2} \right) \right],
\end{eqnarray}
where $\mu^{2}\equiv m^{2}/V_{0}$. Then, we sum-over all time contours $\mathcal{C}$ that begins from the South Pole to a given boundary $\partial a = b$ and $\partial \phi = \psi$, and obtain the wave function
\begin{eqnarray}
\Psi\left[ b, \psi \right] = \int_{\mathcal{C}} \mathcal{D}a \mathcal{D}\phi\;\; e^{-S_{\mathrm{E}}\left[a, \phi\right]}.
\end{eqnarray}
We require the condition that $b$ and $\psi$ should be real values, while $a$ and $\phi$ can be complex-valued functions in general.

We use the steepest-descent method and consider only the on-shell solutions to count the probability from the path-integral. We solve the following classical equations of motion for the Euclidean time direction
\begin{eqnarray}
\label{E3}\ddot{\phi} &=& - 3 \frac{\dot{a}}{a} \dot{\phi} + \mu^{2} \phi,\\
\label{E4}\ddot{a} &=& - \frac{8 \pi}{3}a \left( \dot{\phi}^{2} + 1 + \frac{1}{2}\mu^{2} \phi^{2} \right),
\end{eqnarray}
where the initial conditions of the South Pole is
\begin{eqnarray}
a(0) = 0, \;\;\; \dot{a}(0) = 1, \;\;\; \dot{\phi}(0) = 1.
\end{eqnarray}
Then the on-shell Euclidean action is given by
\begin{eqnarray}
S_{\mathrm{E}} = \frac{4\pi^{2}}{V_{0}} \int d \tau \left( a^{3} + \frac{a^{3}}{2}\mu^{2} \phi^{2} - \frac{3}{8 \pi} a \right).
\end{eqnarray}
When $a$ and $\phi$ approach to real values, the real part of the Euclidean action slowly varies than that of the imaginary part:
\begin{eqnarray}
\left| \nabla_{A} \mathrm{Re} \;S_{\mathrm{E}} \right| \ll \left| \nabla_{A} \mathrm{Im} \;S_{\mathrm{E}} \right|,
\end{eqnarray}
where $A = b, \psi$. If this condition is satisfied, then the probability of the history is well-defined and can be estimated by
\begin{eqnarray}
P \propto e^{-2 \mathrm{Re} \;S_{\mathrm{E}}}.
\end{eqnarray}
Here, we introduce the factor two because the integration is up to the half of the manifold; in any case, the result should be consistent with that of the previous section.

\begin{figure}
\begin{center}
\includegraphics[scale=0.4]{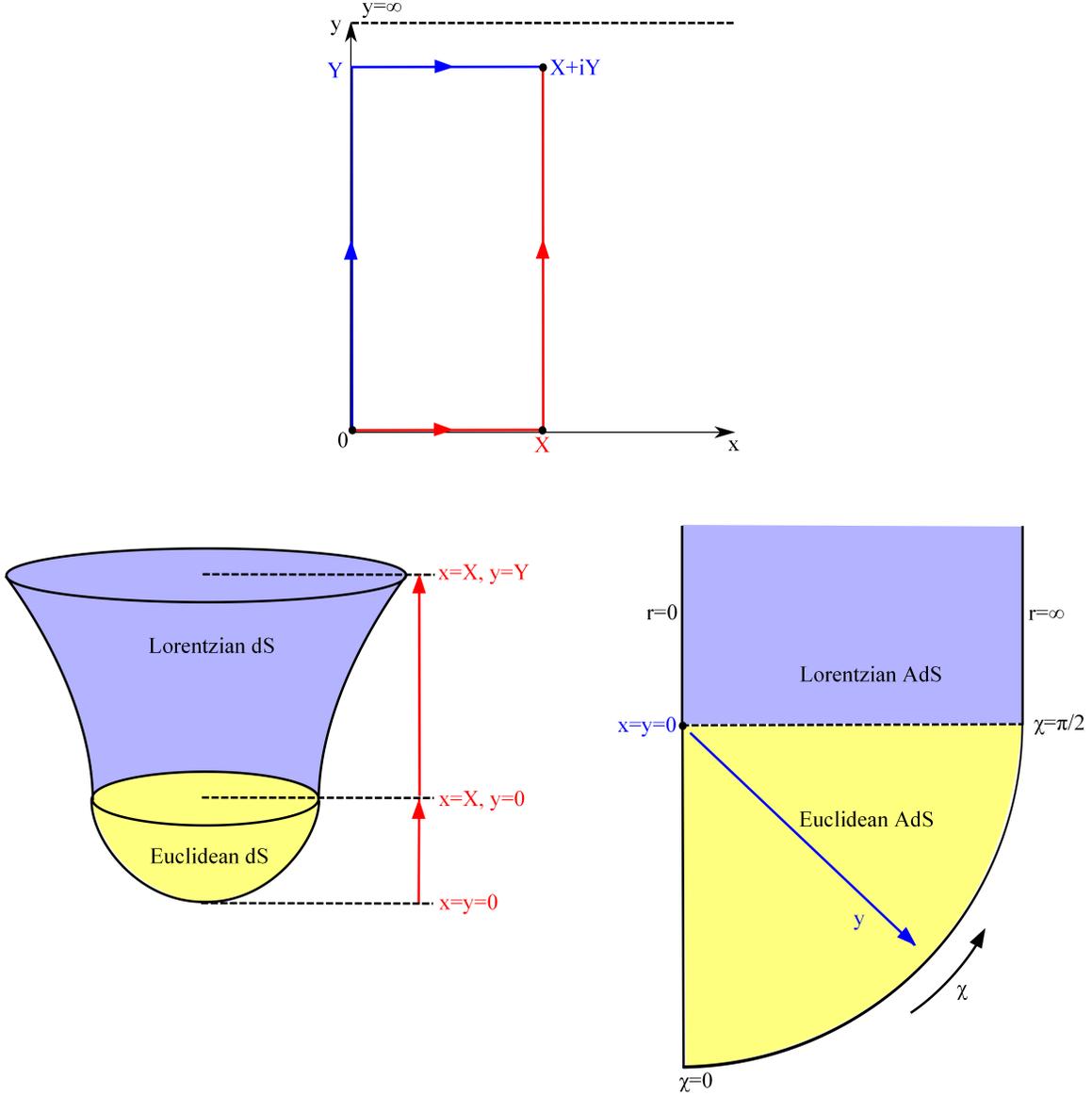}
\caption{\label{fig:path}Upper: Two possible contours. Lower left: The red contour can be interpreted as a homogeneous dS space. Lower right: the blue contour can be interpreted as an inhomogeneous AdS space.}
\end{center}
\end{figure}

\subsection{Choice of time contours}

For a time contour $\mathcal{C}$, if the end point satisfies the classicality conditions, then the probability is well-defined and it does not depend on the choice of $\mathcal{C}$. We regard that a contour is labeled by the time parameter $\tau = x + iy$, where $x$ and $y$ are real. We label the South Pole by $\tau = 0$ and the end point by $\tau = X + iY$. We illustrate two different ways and their analytic continuations to the Lorentzian manifold (FIG.~\ref{fig:path}, see also \cite{Lee:2012qv}). Regarding the choice of different contours, see also \cite{Battarra:2014xoa}.

\subsubsection{Homogeneous dS}

First, we begin from $x = y = 0$ and follow the real axis of $\tau$ until $\tau = X$; then move $\tau = X + iy$ until $y=Y$ (following red contour in FIG.~\ref{fig:path}). Then, the Wick rotation to $y$ direction naturally makes that the instanton follows along the Lorentzian signatures. Therefore, this explains the creation of homogeneous dS universe (lower left of FIG.~\ref{fig:path}). If $\mu < \sqrt{6\pi}$ (this is the same as the BF bound in AdS), then there is a continuous spectrum of the classicalized instantons \cite{Hartle:2007gi} and the probability $P$ will be approximately
\begin{eqnarray}
\log P \simeq \frac{3}{8V_{0}} \left( 1 + \frac{1}{2} \mu^{2} \left| \phi(0) \right|^{2} \right)^{-1},
\end{eqnarray}
where the initial condition of the scalar field is $\phi(0) = \left| \phi(0) \right| e^{i\theta}$ with real valued constants $\left| \phi(0) \right|$ and $\theta$. (In fact, to satisfy the classicality, we have to tune $X$ and $\theta$; hence, the only dynamical degrees of freedom of $P$ is the one dimensional field modulus $\left| \phi(0) \right|$.)

\subsubsection{Inhomogeneous AdS}

Second, we begin from $x = y = 0$ and follow the imaginary axis of $\tau$ until $\tau = iY$; then move through $\tau = x + iY$ until $x = X$ (following blue contour in FIG.~\ref{fig:path}). Note that we should impose
\begin{eqnarray}
\frac{\partial a}{\partial x} = \frac{\partial a}{i \partial y},
\end{eqnarray}
and hence $a_{,y}(x=0,y=0) = -i$. Therefore, along the $x=0$ axis, $a(x=0,y)$ should be dominated by the imaginary parts.

So, to describe the $x=0$ axis, it is convenient to rewrite $a = -i\alpha$. Then, by using $d\tau = i dy$,
\begin{eqnarray}
S_{\mathrm{E}} = \frac{2 \pi^{2}}{V_{0}} \int dy \left[ -\frac{3}{8\pi} \left( \alpha \dot{\alpha}^{2} + \alpha \right) + \frac{1}{2}\alpha^{3} \dot{\phi}^{2} + \alpha^{3} \left ( -1 - \frac{1}{2}\mu^{2} \phi^{2} \right) \right],
\end{eqnarray}
where $\dot{~}$ is now the differentiation for $y$. Therefore, this means that along the $x=0$ line, the solution behaves as the Euclidean AdS with the potential
\begin{eqnarray}
U(\phi) = U_{0} - \frac{1}{2} m^{2} \phi^{2} = - V(\phi),
\end{eqnarray}
where $U_{0} = - V_{0}<0$.

If $Y \rightarrow \infty$, then $\alpha(x=0,y)$ forms a complete Euclidean AdS and the contribution for $\tau = x + iY$ only locates on the boundary of the Euclidean AdS. For the $x=0$ slice, we can Wick rotate to the Lorentzian signature by $\chi = \pi/2 + it$. This explains a creation of an inhomogeneous AdS universe (lower right of FIG.~\ref{fig:path}). Again, the probability is the same as the previous subsection, so that
\begin{eqnarray}
\log P \simeq \frac{3}{8\left|U_{0}\right|} \left( 1 + \frac{1}{2} \mu^{2} \left| \phi(0) \right|^{2} \right)^{-1}.\nonumber
\end{eqnarray}

\begin{figure}
\begin{center}
\includegraphics[scale=0.5]{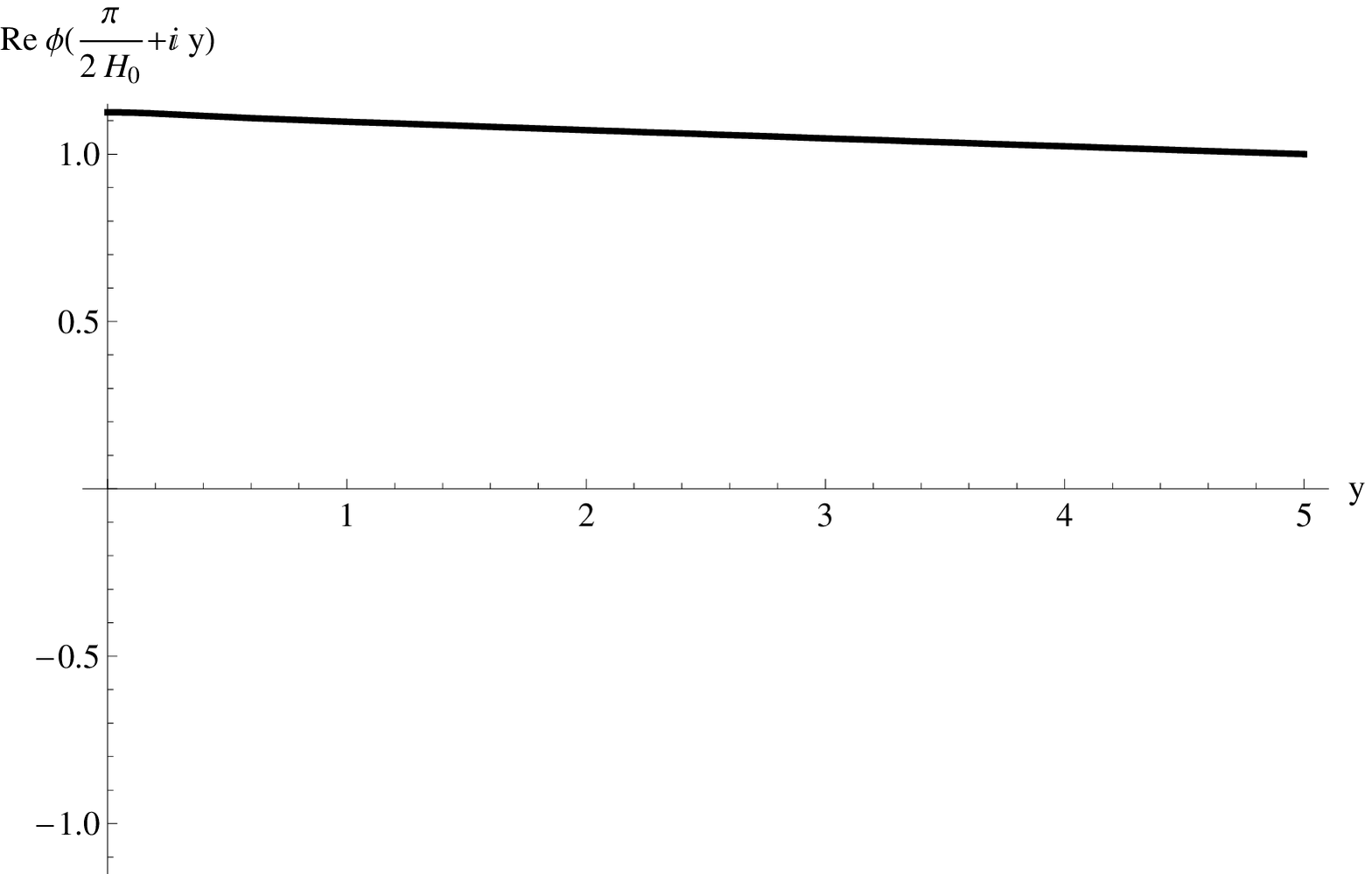}
\includegraphics[scale=0.5]{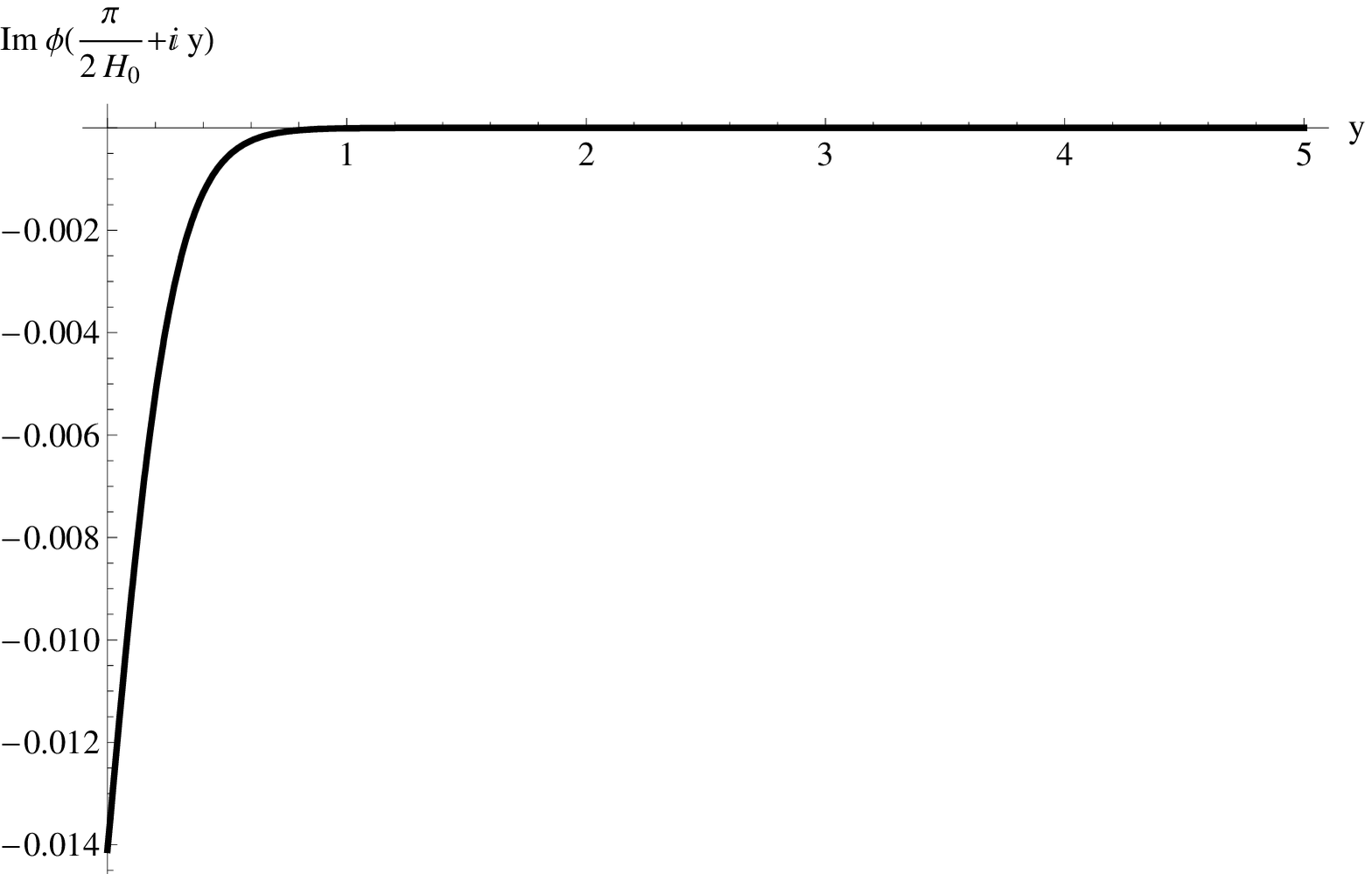}
\includegraphics[scale=0.5]{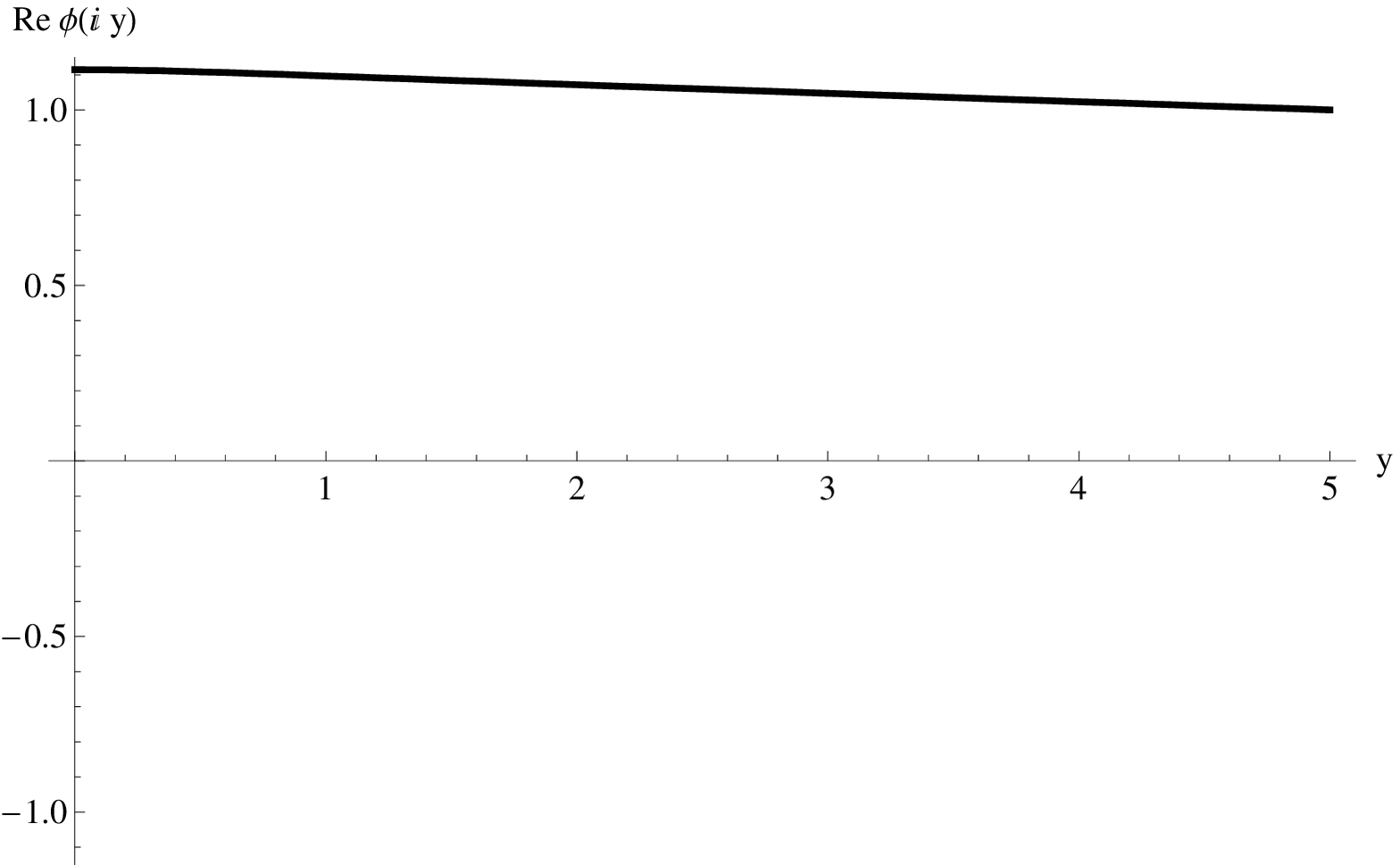}
\includegraphics[scale=0.5]{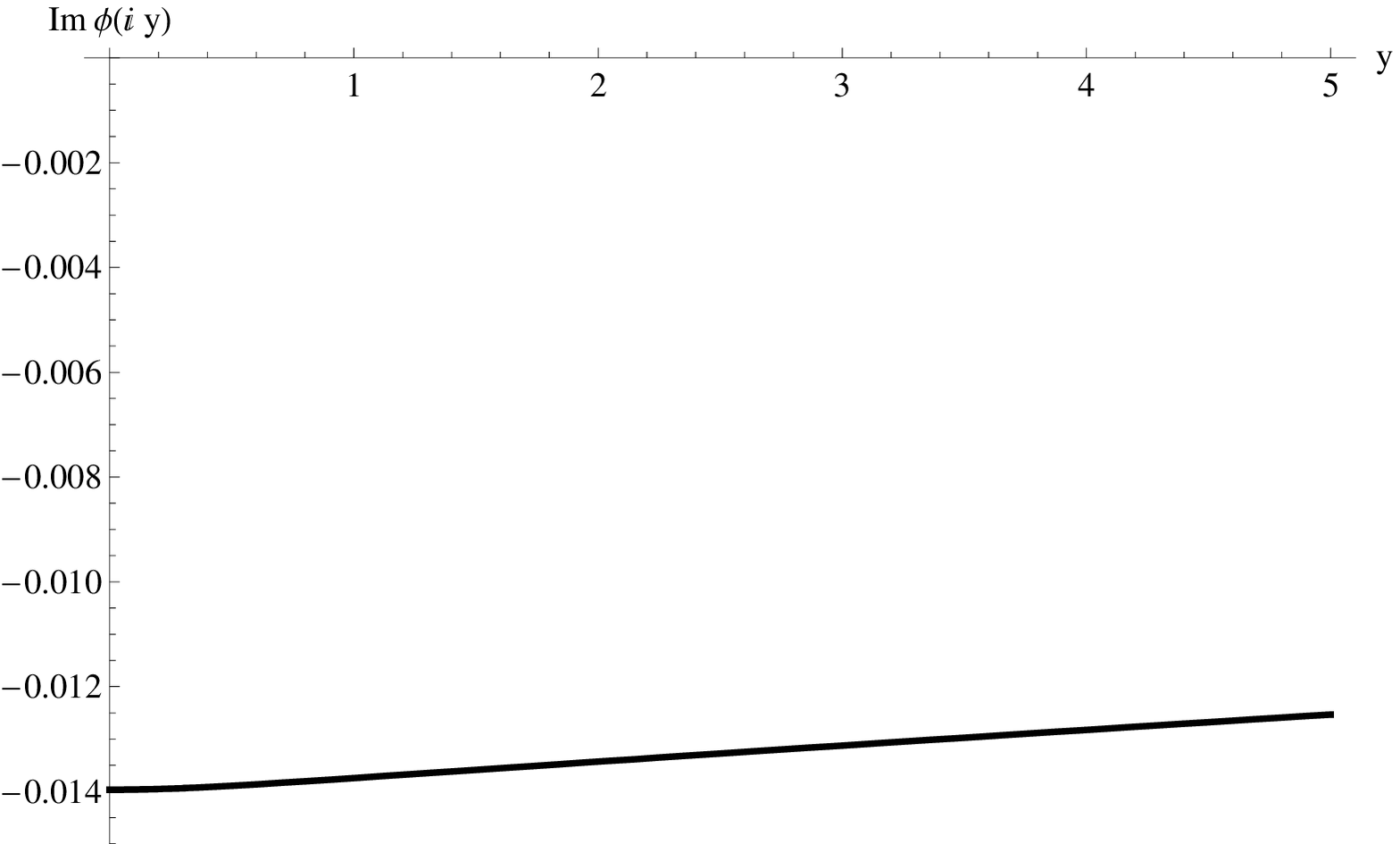}
\caption{\label{fig:reimphi}Perterbative solutions $\phi$ for different contours with the conditions $\mu^{2}=0.2$, $Y=5$, and $\phi(\pi/2H_{0} + iY) = 1$. Upper: $\phi(\pi/2H_{0} + iy)$ for the real part (left) and the imaginary part (right). Lower: $\phi(iy)$ for the real part (left) and the imaginary part (right). The upper is classicalized while the lower is not.}
\end{center}
\end{figure}

\subsubsection{No-boundary regulator}

Now we come back to the instantons of the tachyonic top in AdS space with the potential:
\begin{equation}
U(\phi) = U_{0} - \frac{1}{2} m^{2} \phi^{2}.\nonumber
\end{equation}
If we believe that the Euclidean action for an AdS background can be regularized by the previous subsection, then we can obtain a non-zero probability even for an AdS background.

Then the regularized Euclidean action for Euclidean AdS becomes
\begin{eqnarray}
\mathrm{Re} \;S_{\mathrm{E,reg}} &=& \mathrm{Re} \; \left[ \lim_{Y\rightarrow \infty} \left[ \int_{0}^{Y} \left. L_{\mathrm{E}} \right|_{x=0} dy + \int_{0}^{X} \left. L_{\mathrm{E}} \right|_{y=Y} dx\right] \right]\\
&=& (\mathrm{Euclidean\;action\;for\;AdS}) + (\mathrm{boundary\;counter\;term \; at\;} Y=\infty)\\
&=& (\mathrm{Euclidean\;action\;for\;dS\;} (U\rightarrow V=-U)),
\end{eqnarray}
and hence the decay rate is
\begin{eqnarray}
\log P &=& 2 \left( \mathrm{Re} \;S_{\mathrm{E,reg}} \left(\mathrm{solution}\right) - \mathrm{Re} \;S_{\mathrm{E,reg}} \left(\mathrm{background}\right) \right)\\
&\simeq& \frac{3}{8\left|U_{0}\right|} \left[ \left( 1 + \frac{1}{2} \mu^{2} \left| \phi(0) \right|^{2} \right)^{-1} - 1 \right].
\end{eqnarray}

\subsection{Classicality and interpretations}

The idea of the no-boundary regulator was introduced by Hartle and Hertog \cite{Hertog:2011ky}. However, in the previous paper, they did not discuss the detailed physical meanings of AdS instantons. Here, we explicitly show the solutions and check their classicality. These will give physical meanings.

\subsubsection{Classicality: scalar field}

Let us fix the background by $a = H_{0}^{-1} \sin H_{0} \tau$ and consider the case when the field values are near the local minimum, where $H_{0}^{2} = 8\pi/3$. Then, the equation becomes
\begin{eqnarray}
\ddot{\phi} + 3 H_{0} \cot H_{0} \tau \dot{\phi} - \mu^{2} \phi = 0.
\end{eqnarray}
This is a linear differential equation. $G(\tau)$ is the solution that satisfies $\dot{G}(0) = 0$ \cite{Hartle:2007gi}, if $G(\tau) = F(a_{+},a_{-};2;z(\tau))$, where $F(a_{+},a_{-};2;z)$ is a hypergeometric function and
\begin{eqnarray}
a_{\pm} &=& \frac{3}{2} \left( 1 \pm \sqrt{1 - \frac{\mu^{2}}{6\pi}} \right),\\
z(\tau) &=& \frac{1 - \cos H_{0}\tau}{2}.
\end{eqnarray}

For the classicalization at $\tau = \pi/2H_{0} + iY$, the phase angle
\begin{eqnarray}
\arctan \frac{\mathrm{Im} \;G(\pi/2H_{0} + iY)}{\mathrm{Re} \;G(\pi/2H_{0} + iY)}
\end{eqnarray}
should be tuned. Note that if $\mu^{2} < 6\pi$, then along the $\tau = \pi/2H_{0} + iy$ contour, the equation becomes
\begin{eqnarray}
\ddot{\phi} + 3 H_{0} \dot{\phi} + \mu^{2} \phi = 0,\nonumber
\end{eqnarray}
and hence $\mathrm{Re}\; \phi$ and $\mathrm{Im}\; \phi$ both behave as over-damped solutions. This makes the parameter $\theta$ to be constant and does not depend on time. Then
\begin{eqnarray}
\phi(\tau) = \psi \frac{G(\tau)}{G(\pi/2H_{0} + iY)}
\end{eqnarray}
is the solution with the classicality at $\tau = \pi/2H_{0} + iY$ and $\phi(\pi/2H_{0} + iY) = \psi$ (Upper of FIG.~\ref{fig:reimphi}).

On the other hand, along the $\tau = 0 + iy$ contour, the phase angle becomes (Lower of FIG.~\ref{fig:reimphi})
\begin{eqnarray}
\frac{\mathrm{Im}\; \phi(iy)}{\mathrm{Re}\; \phi(iy)} \simeq \frac{\mathrm{Im}\; \phi(0)}{\mathrm{Re}\; \phi(0)} \simeq \mathrm{const}.
\end{eqnarray}
Therefore, along the contour of $\tau = iy$, the scalar field never be classicalized.

This implies an interesting implication. If we apply the no-boundary regulator, then there exist a continuous spectrum of solutions around the tachyonic top (similar as dS with a massive potential). However, to apply the no-boundary regulator, the field values should be complex-valued. This is also true for dS case. On the other hand, one difference is as follows: for the dS case, after the Wick rotation, the solution approaches to be classicalized and hence the imaginary parts will soon disappear to the Lorentzian observers, while for the AdS case, the imaginary field values cannot be realized even after the Wick rotation.

\subsubsection{Classicality: metric back-reactions}

\begin{figure}
\begin{center}
\includegraphics[scale=0.28]{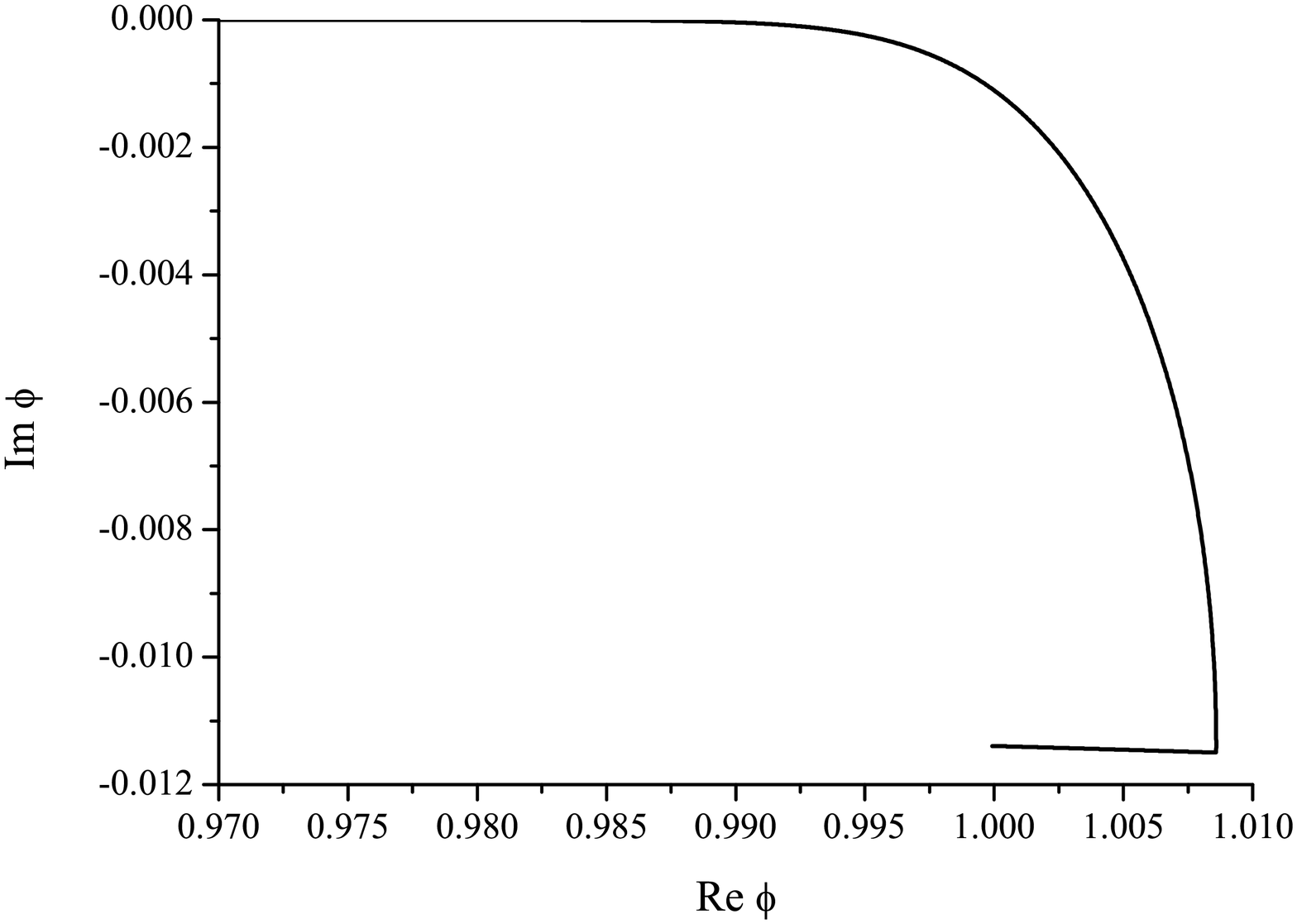}
\includegraphics[scale=0.28]{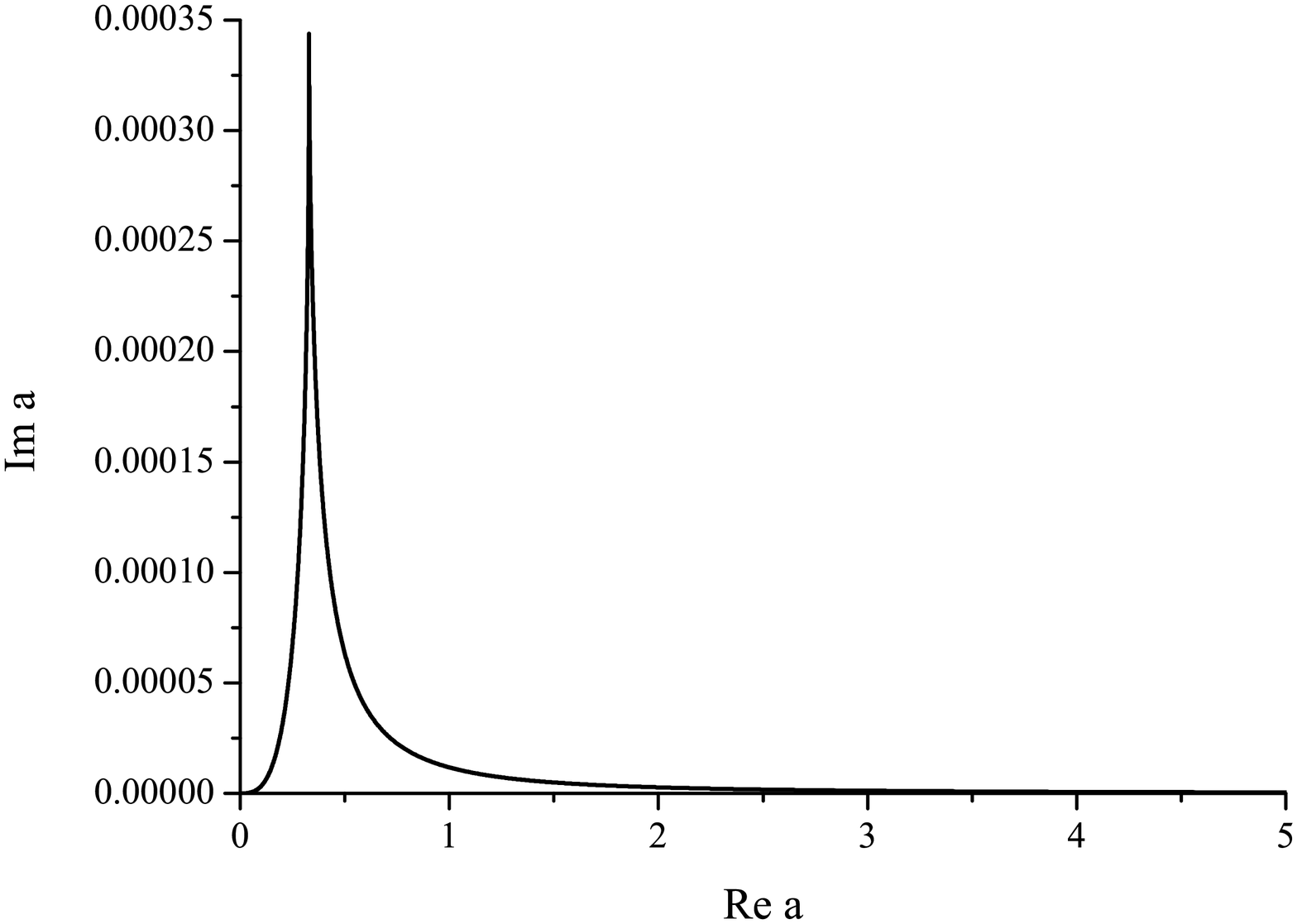}
\caption{\label{fig:exm}Example of a classicalized instanton for $|\phi(0)|=1$ and $\mu^{2}=0.2$, where left is $\phi$ and right is $a$. Around the cusp, the contour rotates from the Euclidean time to the Lorentzian time. The solution is classicalized for $\theta = -0.01139376$ and $X = 0.51732875$.}
\end{center}
\end{figure}

\begin{figure}
\begin{center}
\includegraphics[scale=0.6]{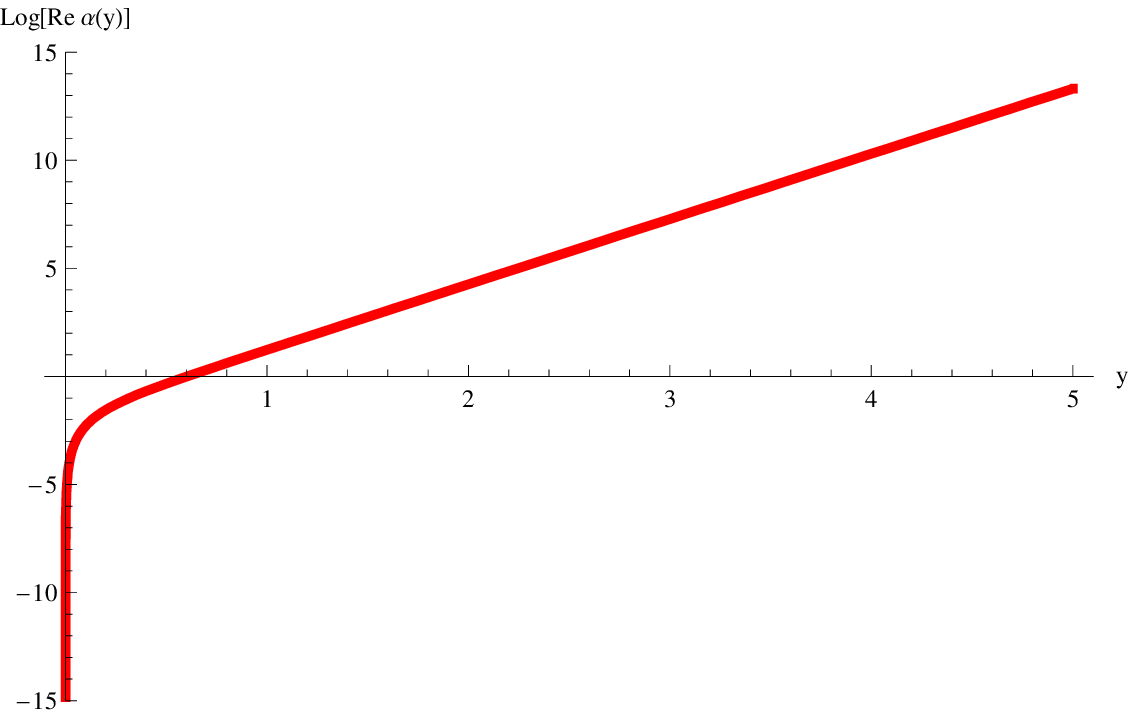}
\includegraphics[scale=0.6]{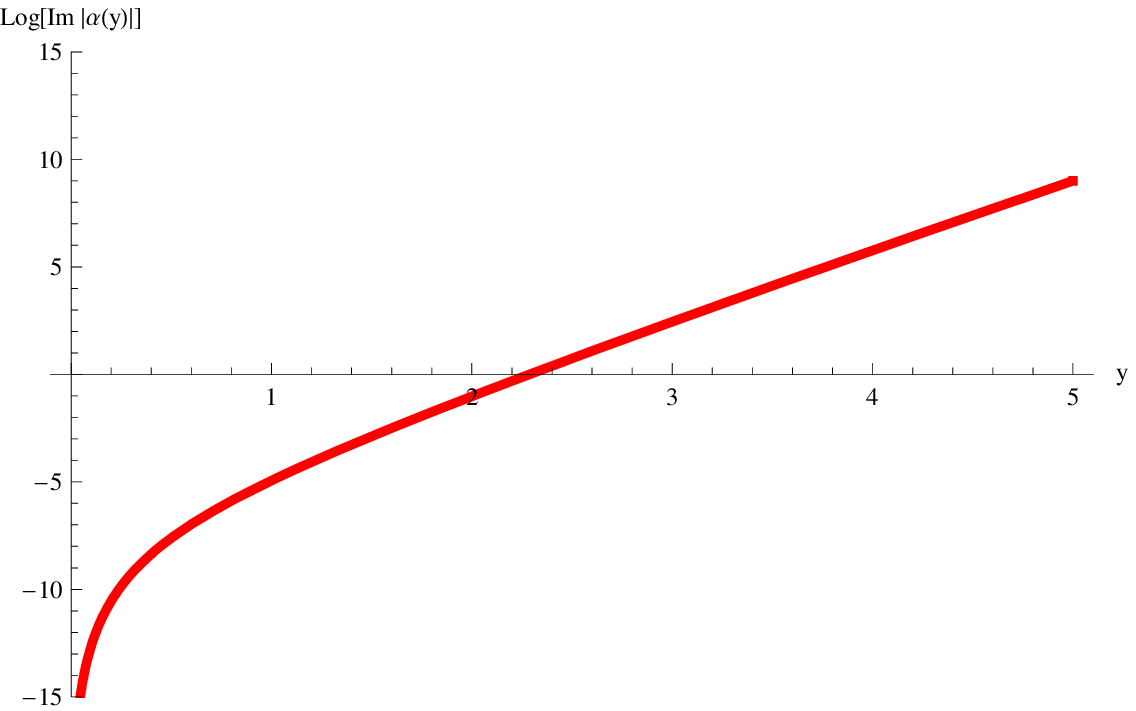}
\includegraphics[scale=0.6]{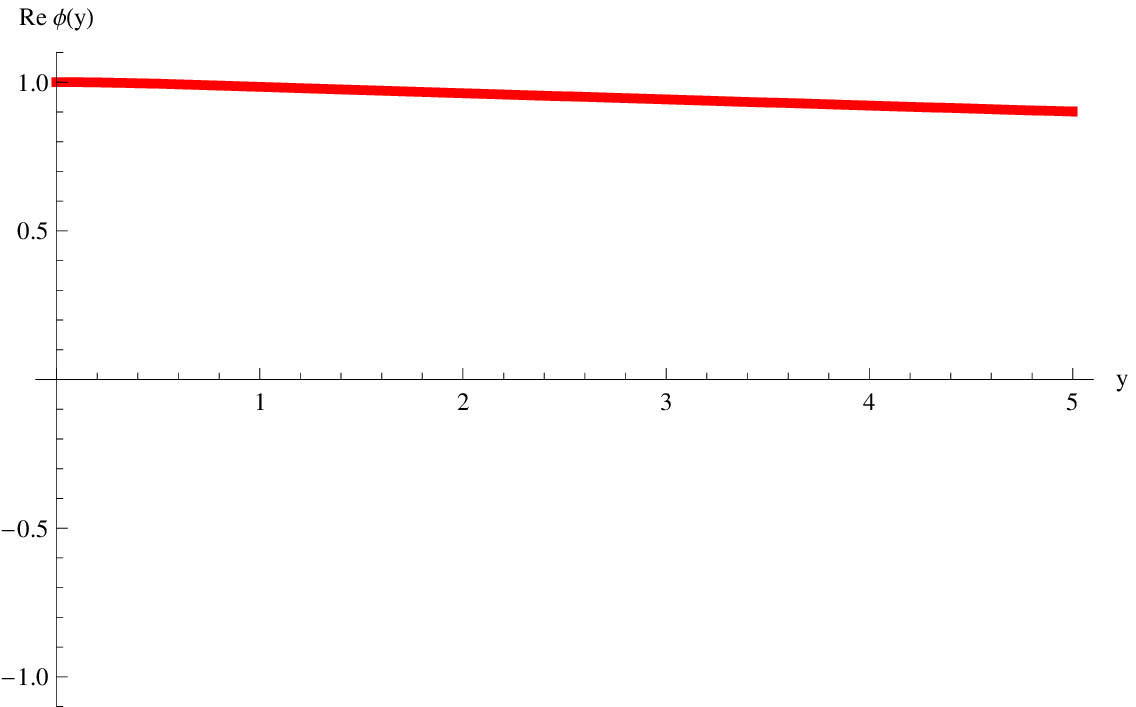}
\includegraphics[scale=0.6]{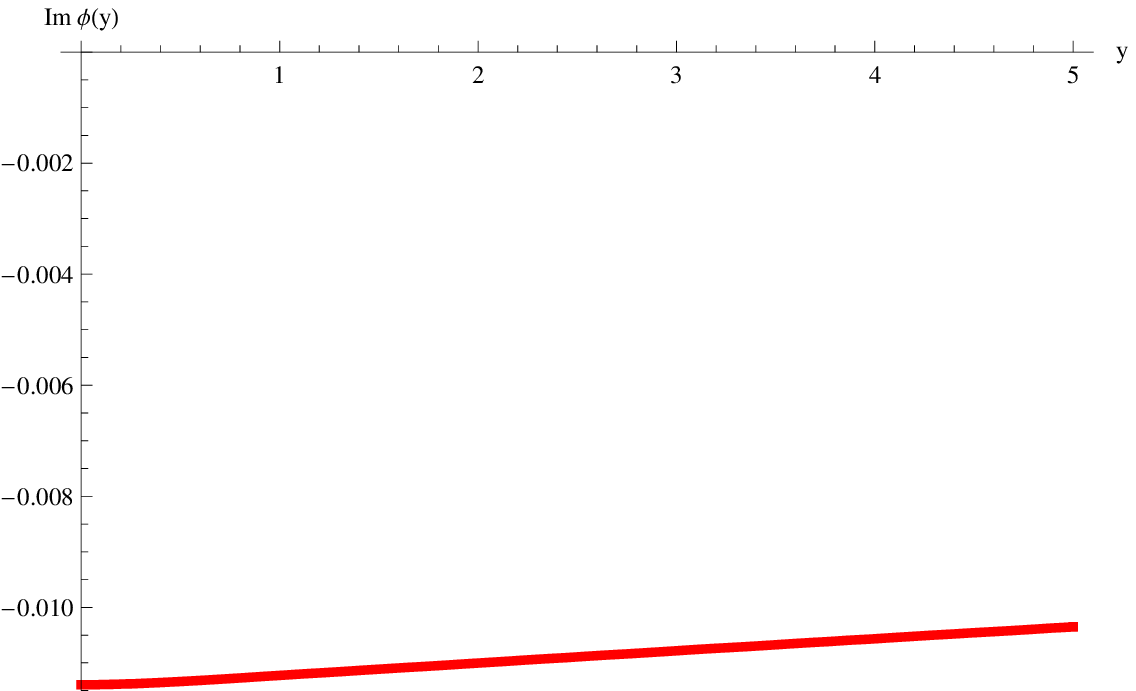}
\caption{\label{fig:num}Solution along the contour $x=0$, with $|\phi(0)|=1$ and $\mu^{2}=0.2$. Upper are log of the real part (left) and the imaginary part (right) of $\alpha$, where the real part is always positive and the imaginary part is always negative. Lower are the real part (left) and the imaginary part (right) of $\phi$.}
\end{center}
\end{figure}

We can go further from the analytic calculations. For the same setting, we can calculate by a numerical way. FIG.~\ref{fig:exm} is one example of the realized instanton by tuning $\theta$ and $X$. We can see that as time goes on, the imaginary part of the metric and the scalar field both approach to zero, while the real parts remain reasonable sizes.

By using the same conditions, we can follow the solution along the contour $x=0$ (FIG.~\ref{fig:num}). If we briefly look at the scalar field (lower left and right), then this is consistent with the analytic results of FIG.~\ref{fig:reimphi} (lower left and right). The new information is on the metric function $\alpha$ along $x=0$. Surely, the imaginary part also increases as $y$ increases. It is fair to say that there is an exponential hierarchy between the real part of $\alpha$ and the imaginary part of $\alpha$. Therefore, the classical metric approximation seems not too bad; in addition, the imaginary part of the scalar field will be vanished for the large $y$ (and hence large $r$) limit. On the other hand, the classicality condition for the metric is no more true for the region near the center.

\subsubsection{Interpretations}

Therefore, we interpret the instantons of tachyonic top in AdS as follows:
\begin{enumerate}
\item For only real-valued instantons, the decay rate becomes zero (unless the potential is finely tuned). However, if we allow the complexification of instantons, then by using the no-boundary regulator, we can obtain a non-zero probability.
\item The classicality is obtained for the point $\tau = X + iY$, where we impose the $Y \rightarrow \infty$ limit. Therefore, the classicality \textit{cannot} be imposed for the part of contour $x = 0$, in the sense that the imaginary part of $\phi$ and $\alpha$ remain significantly after the Wick rotation.
\item Although the tunneling is possible with a non-zero probability, this does not necessarily mean that the tachyonic potential is unstable via quantum processes. These continuous spectrum of instantons around the tachyonic top will be interpreted by small perturbations after the Wick rotation (of course, there can exist instantons with large amplitudes but these probabilities will be highly suppressed; so, for dominant instantons, these will be interpreted by small perturbations). Since the mass scale is less than the BF bound, the local maximum will not be destabilized. This is similar to the case of dS; these come from quantum/thermal fluctuations or excitations (not the bounce that destabilizes a local minimum) and quantum/thermal fluctuations \textit{can} destabilize but \textit{do not necessarily} mean the destabilization.
\end{enumerate}

The no-boundary regulator effectively cuts the outside the radius of AdS $r > l \sim |\Lambda|^{-2}$, where $\Lambda < 0$ is the cosmological constant of AdS. This is physically the same as the case of dS; the Euclidean instanton cuts outside the horizon $r > l \sim |\Lambda|^{-2}$, where $\Lambda > 0$ is the cosmological constant of dS. Both of dS and AdS give non-vanishing $\langle \phi^{2} \rangle > 0$.

The positive dispersion can be interpreted by the uncertainty of the scalar field: $(\Delta \phi)^{2} = \langle \phi^{2} \rangle$ (since $\langle \phi \rangle = 0$). Regarding this uncertainty of the field (as well as the metric, since the field will affect to the metric), interpretations should be different.
\begin{enumerate}
\item For dS, each instanton is classicalized independently. Therefore, even though $\langle \phi^{2} \rangle$ is non-zero, one observer only sees one special history with the initial condition $|\phi(0)|$. The only superspace observer (omnipotent observer) can see the dispersion of quantum states, while we (semi-classical observers) cannot be such a superspace observer.
\item For AdS, each instanton is not classicalized. This implies that there is no classical observer who can distinguish the independent instantons. This further implies that the asymptotic observer (even semi-classical observer) may notice the uncertainty of the scalar field, i.e., the fuzzy (orbital-like) field/metric distribution via quantum fluctuations.
\end{enumerate}
This uncertainty can be seen by only an AdS boundary observer and not by a dS observer. Of course, the decisive interpretation on the unclassicalized solution is still difficult. This means that we are dealing with a genuinely quantum gravitational phenomenon. So, further investigations are necessary.

\section{\label{sec:dis}Discussion}

In this paper, we revisited the quantum stability issue around the tachyonic top in AdS. If the mass scale is smaller than the BF bound, then it is perturbatively stable. We introduced the no-boundary regulator, and this makes a continuous instanton spectrum around the tachyonic top, while these instantons may not destabilize the equilibrium; rather these mean quantum/thermal fluctuations.

One interesting observation is on the interpretation of the positive dispersion of the scalar field. For the dS case, each instanton is classicalized and hence a semi-classical observer cannot see the dispersion. On the other hand, for the AdS case, each instanton cannot be classicalized and hence the boundary semi-classical observer may see the dispersion of the scalar field.

Then what does it look like? What is the thing that the asymptotic AdS observer sees? This goes beyond the scope of this paper. However, what we can say is that due to the quantum fluctuation of the scalar field, perhaps the scalar field is uncertain and the metric is not classical; therefore, due to quantum fluctuations, general relativity seems to be effectively broken, in the sense that when the metric is not classicalized, the metric does not satisfy classical equations of motion, while general relativity is a kind of classical equations of motion. To summarize, if we believe that AdS is unitary, then it seems that such unitary quantum effects may induce the violation of classical equations of motion.

Of course, it is fair to say that this is a brave interpretation at this stage. However, if it is the case, can this resolve the contradiction between unitary quantum field theory and general relativity in the context of black hole complementarity controversy \cite{Susskind:1993if,Yeom:2008qw,Almheiri:2012rt}? In addition, the non-zero dispersion may be interpreted in the context of the uncertainty principle \cite{Chen:2014bva} though it requires further justifications. Still this paper cannot cover black hole physics, but it gives some intuitions of AdS. Further study of Euclidean quantum gravity and the information loss problem \cite{Sasaki:2014spa} can be an interesting topic for future study.

\newpage

\section*{Acknowledgment}

This work was supported by the National Research Foundation of Korea (NRF) grant funded by the Korea government (MSIP, No.~2014R1A2A1A01002306).
DY was supported by the JSPS Grant-in-Aid for Scientific Research (A) No.~21244033. DY was also supported by Leung Center for Cosmology and Particle Astrophysics (LeCosPA) of National Taiwan University (103R4000).
WL was supported by Basic Science Research Program through the National Research Foundation of Korea (NRF) funded by the Ministry of Education, Science and Technology (2012R1A1A2043908).

\end{document}